\documentclass[twocolumn,showpacs,twoside,10pt,superscriptaddress,prl]{revtex4}

\usepackage{graphicx}
\usepackage{epsfig}
\usepackage{epsf}
\usepackage{amssymb}
\usepackage{amsmath}
\usepackage{amsthm}
\usepackage[colorlinks=true,dvipdfm]{hyperref}

\begin{document}

\title{Complete Adiabatic Quantum Search in Unsorted Databases}
\author{Nanyang Xu, Xinhua Peng, Mingjun Shi, Jiangfeng Du$^{\ast}$\\
\normalsize{Hefei National Laboratory for Physical Sciences at
Microscale and Department of Modern
Physics,University of Science and Technology of China, Hefei, Anhui 230026, People's Republic of China}\\
\normalsize{$^\ast$To whom correspondence should be addressed; E-mail:  djf@ustc.edu.cn}}

\begin{abstract}
We propose a new adiabatic algorithm for the unsorted database
search problem. This algorithm saves two thirds of qubits than
Grover's algorithm in realizations. Meanwhile, we analyze the time
complexity of the algorithm by both perturbative method and
numerical simulation. The results show it provides a better speedup
than the previous adiabatic search algorithm.

\end{abstract}


\maketitle
%\section{Introduction}
Quantum computation is a promising way to solve classical hard
problems. Several quantum algorithms have been designed to perform
classical algorithms with remarkable speedups. The most useful one
among these is Grover's algorithm\cite{Grover_search} which concerns
the problem of searching for a required item in a unsorted database.
One common example for this unsorted database search is to find a
person's name in a phone book (the items are sorted by names) with
only knowing his phone number. Classically, the only way to achieve
this is brute-force search\cite{Grover_search,Ju_2007} which
requires an average of $ \frac{N}{2}$ quires for $N$ entries in the
phone book. However, if the information is stored in a quantum
database, to find the right name with Grover's algorithm costs only
a time of order $\sqrt{N}$, providing a quadratic speedup.

The main process of Grover's algorithm is, swinging the $index$
qubits from an initial uniform state to approach the solution state.
The information of the database is not explicitly accessed in the
processing of search. Instead, an Oracle is supposed to know all the
information in the database and act properly towards a input state
depending on whether it denotes the solution\cite{QCQI}. Early
experiments\cite{Chuang_1998search, Dodd_2003,
Vandersypen_3bitsearch, Brickman_2005,Long_2001} of Grover's
algorithm constructed this Oracle from a marked state as a
functional analog instead of querying the database. For a complete
solution of search problem, Kim and coworkers proposed a new
approach to realize the Oracle on a quantum database and implemented
it in experiment\cite{Kim_2002}. In this complete approach of
Grover's algorithm, extra qubits are used to store the database.
Similar method for construction of Grover's Oracle was also
theoretically discussed later \cite{Ju_2007}.

While Grover's algorithm is presented in the standard circuit
model(\textit{i.e.}, using a sequence of discrete quantum gates), a
new model of quantum computation shows up where the state of quantum
computer evolves continuously and adiabatically under a certain
time-dependent Hamiltonian. This new adiabatic model was soon
applied to the database search problem\cite{Farhi_2000} and the
original adiabatic search algorithm was proved to have a time
complexity of order $N$, which is the same performance as classical
algorithms. More recently, Roland and Cerf\cite{Roland_2002}
recovered the advantage of adiabatic search to order $\sqrt{N}$ (the
same with Grover's algorithm), by performing the adiabatic evolution
locally. However this adiabatic search algorithm construct the
Hamiltonian from a marked state instead of referring to the
database, thus it is not a complete search algorithm. And in order
to discriminate it from our algorithm, we call it \emph{the
marked-state adiabatic search}(MSAS) algorithm in the following
paragraphs.

In this letter, we apply the quantum adiabatic computation to
unsorted database search problem again and present a new quantum
search algorithm. We will put forward a new method to represent the
database. By this method the algorithm contains no Oracles and saves
$\frac{2}{3}$ of qubits than the complete approach of Grover's
algorithm\cite{Kim_2002}. We also analyze the time complexity by
both perturbation method and numerical simulation. The results show
it provide a higher speedup than the MSAS algorithm.

%\section{Adiabatic Quantum Computation}

As a new quantum computation model, adiabatic algorithm was brought
out by Farhi \textit{et al.}\cite{Farhi} and soon became a rapidly
expanding field. The idea of this new computation model is to
prepare a system in the ground state of a simple initial
Hamiltonian, then slowly switch the simple Hamiltonian to a complex
Hamiltonian whose ground state encodes the solution to the problem
of interest. According to Adiabatic Theorem, the the system stays in
the ground state of the instantaneous Hamiltonian if we perform the
evolution slowly enough. So finally the state describes the solution
to the problem. The time-dependent system Hamiltonian is
\begin{equation}
\label{ht}\\
 H(t)=[1-s(t)]H_{i}+s(t)H_{p},
\end{equation}
where $H_{i}$ is the initial Hamiltonian and $H_{p}$ is the problem
Hamiltonian which encodes the solution, and the monotonic function
$s(t)$ fulfills $s(0) = 0$ and $s(T)=1$.

%\section{Non-Oracle quantum search algorithm}
Here let's focus on the unsorted database search problem. To be
simplified, the database is a list of $(i,v_i)$ pairs and sorted by
$i$ where $i$ denotes $index$ and $v_i$ is $value$. Both $i$ and
$v_i$ are $n$-bit binary codes thus the database contained $N=2^n$
items. The \emph{"unsorted"} property of the database refers to the
field $value$ not $index$. The unsorted database search problem here
is looking for the corresponding $index$ $i$ for a given target
$value$ ${t}$. And we assume that there's only one solution in the
database for each search. Next we will describe the process to find
the right $i$ which connects to the target ${t}$.

The essential part of an adiabatic search algorithm is how to encode
the solution in the ground state of problem Hamiltonian. For
example, the MSAS algorithm constructs the problem Hamiltonian as
$H_p=1-|m\rangle\langle m|$ where $|m\rangle$ is exactly the
solution state. Thus it is not a complete database search. Obviously
for a complete search, the information in database should be
represented in quantum forms. Taking the complete approach of
Grover's algorithm as an example\cite{Kim_2002,Ju_2007}, the
database is represented in an operator which satisfies $
U_{f}|i\rangle|0\rangle=|i\rangle|v_{i}\rangle$. $U_f$ generates the
entanglement of qubits to denote the relation between $i$ and $v_i$,
thus both the fields are represented by qubits.

In the present algorithm, however, not both the fields are
represented by qubits. We define a database operator as
\begin{equation}
\label{dop} \mathcal{D}=\sum_{i=0}^{N-1} v_{i}|i\rangle \langle i|.
\end{equation}
Clearly in this approach, the $index$ is represented by qubits while
the $value$ is store in the strength of interactions. So no extra
qubits are needed for the database. The operator $\mathcal{D}$
contains all the information in the database. Thus, we can construct
the problem Hamiltonian from $\mathcal{D}$ as
\begin{eqnarray}
\label{hp1} H_{p} &=& (\mathcal{D} - {t})^2,
\end{eqnarray}
where ${t}$ is the target $value$ which we are looking for.

To test the validity of $H_p$, we will examine its ground state. To
this end, we can write $H_p$ as $ \sum_{i=0}^{N-1}
(v_{i}-{t})^2|i\rangle \langle i|$. From this form, each diagonal
element of $H_p$ is the square of difference between $v_i$ and
${t}$. Thus the ground state will be the solution state $|i\rangle$
where $v_i$ equals to ${t}$.

Of course this construction provides a valid problem Hamiltonian for
the search problem. However, the Hamiltonian in Eq.\eqref{hp1} has a
spectral width exponentially growing with the number of qubits,
which is hard to realize when the database is large. Thus it is only
useful in small-size databases.

To solve this problem, we divide the comparison between $v_i$ and
${t}$ into $n$ sub-comparisons, each of which is performed for a
single bit between them. Thus the database operator should be formed
separately for each bit. For the $j$th bit of $value$ we define the
bit database operator $\mathcal{D}_j$ as
\begin{eqnarray}
\label{hpd1} \mathcal{D}_{j} &=& \sum_{i=0}^{N-1} v_{ij}|i\rangle
\langle i|,
\end{eqnarray}
where $v_{ij}$ is the $j$th bit of $v_{i}$. Similarly with the
operation in Eq.\eqref{hp1}, the problem Hamiltonian for each bit is
\begin{eqnarray}
H_{p}^{j} &=& (\mathcal{D}_{j} - {t}_{j})^{2},
\end{eqnarray}
where ${t}_{j}$ is as well the $j$th bit of ${t}$. Consequently, the
overall problem Hamiltonian is the summarization of all bit problem
Hamiltonians
\begin{eqnarray}
\tilde{H_{p}} &=& \sum_{j=0}^{n-1}{H_{p}^{j}} =
\sum_{j=0}^{n-1}{[\mathcal{D}_{j}(1-{t}_{j}) + {t}_{j}(I - \mathcal{D}_{j})]}\nonumber\\
&=&
\sum_{j=0}^{n-1}{(\mathcal{D}_{j}\bar{t}_{j}+{t}_{j}\bar{\mathcal{D}}_{j})},
\end{eqnarray}
where $\bar{t}_{j}$ is the complementation of binary bit ${t}_{j}$
and $\bar{\mathcal{D}}_j = I - \mathcal{D}_j$.

Also for a test of the validity, we can simplify $\tilde{H_{p}}$ as
\begin{eqnarray}
\label{hp_hd} \tilde{H_{p}} &=& \sum_{i=0}^{N-1}
{h(v_{i},t)}|i\rangle \langle i|,
\end{eqnarray}
where the function $h(v_i,{t})$ is the Hamming distance between
$v_i$ and $t$. Thus the state $|i\rangle$ where $h(v_i,t)=0$ is the
ground state of $\tilde{H_{p}}$ and is also the solution state.
Moreover, the spectrum was successfully bounded in a range from $0$
to $n$.

After the preparation of problem Hamiltonian, we will choose an
initial Hamiltonian $H_{i}$. $H_{i}$ should be chosen to be
noncommutative with $H_{p}$ to avoid crossing of energy
levels\cite{Farhi_2000}. Normally, $H_{i}$ is
\begin{eqnarray}
\label{initH} H_{i} & = & g(\sigma_{x}^{0}+\sigma_{x}^{1}+\cdots +
\sigma_{x}^{n-1}),
\end{eqnarray}
which means the qubits coupling with a magnetic field at the
$x$-direction and the coupling strength is $g$. The ground state of
$H_i$ is
\begin{eqnarray}
\label{initS} |\psi_{0}\rangle & =&
\frac{1}{\sqrt{N}}\sum_{j=0}^{N-1}(-1)^{b(j)}|j\rangle,
 \label{Int_state}
\end{eqnarray}
where $b(j)$ is the Hamming distance between $j$ and $0$.

In the adiabatic evolution, the system Hamiltonian interpolates from
$H_{i}$ to $\tilde{H_{p}}$ (\textit{i.e.,} see Eq \ref{ht}) and the
state of the system evolves according to the Schr\"{o}dinger
equation. If this evolution acts adiabatically, the system will
always stay on the instantaneous ground state of $H(t)$ and in the
end the solution of our problem will show up.

An explicit application is necessary for a clear understanding. Here
we perform a 3-bit unsorted database search for example. We randomly
generate a database in a list as $\{6,3,5,0,4,1,7,2\}$. The position
of each $value$ in the list refers to the $index$ which ranges from
0 to 7. For convenience we rewrite the database as binary codes
which is \{110,011,101,000,100,001,111,010\}. Because the database
operators and problem Hamiltonian are diagonal, they are expressed
by only the diagonal elements.
\begin{eqnarray}
\mathcal{D}_0 &=& diag\{0,1,1,0,0,1,1,0\}\nonumber\\
\mathcal{D}_1 &=& diag\{1,1,0,0,0,0,1,1\}\nonumber\\
\mathcal{D}_2 &=& diag\{1,0,1,0,1,0,1,0\}.
\end{eqnarray}

After building the quantum database, the problem Hamiltonian can be
constructed for each search task. For example, if we want to find
the position of the $value$ 5 which is 101 in binary, the problem
Hamiltonian is

\begin{eqnarray}
\label{hp_example}
\tilde{H_{p}} &=& \bar{\mathcal{D}}_{0}
+\mathcal{D}_1 +
\bar{\mathcal{D}}_{2}\nonumber\\
&=& diag\{2,2,0,2,1,1,1,3\}.
\end{eqnarray}

\begin{figure}[t]
\begin{center}
\includegraphics[width= 1\columnwidth]{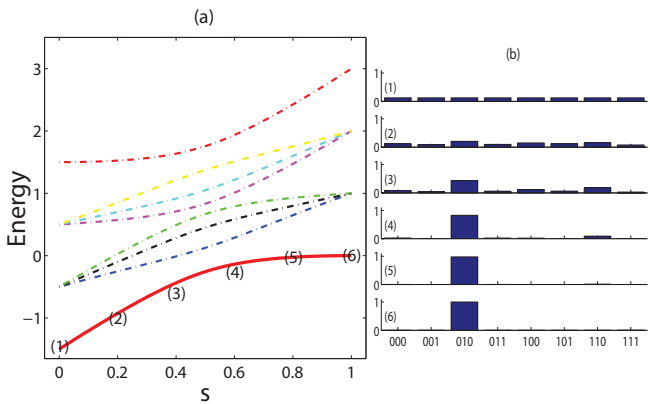}
\end{center}
\caption{Process of the adiabatic evolution to search for $value$ 5
in the mentioned database. (a)The instantaneous eigenvalues of the
system Hamiltonian as a function of $s$. The solid line represented
the energy level of ground state. (b)Occupation probabilities of the
system for the computational basis during the adiabatic evolution in
the numerical simulation. The system starts from a uniform state and
evolves to the solution state $|010\rangle$ which shows $value$ 5 is
on $index$ 2. The parameter $g$ of $H_i$ in this example is 0.5.}
\label{evolution}
\end{figure}

To perform the adiabatic evolution, we initially prepare the system
on the state of Eq.\eqref{initS}, and adiabatically switch the
system Hamiltonian from a initial Hamiltonian in Eq.\eqref{initH} to
the problem Hamiltonian in Eq.\eqref{hp_example}. Finally the system
will on the ground state of $\tilde{H_{p}}$ which is the state
$|2\rangle$. After measurement, we can get the knowledge that
$value$ 5 is on position 2. Fig.\ref{evolution} shows the process of
the adiabatic evolution for this search.

For the practical usefulness of an algorithm, the occupied amount of
resources is an important aspect. Without doubts, the number of
qubits needed for our algorithm equals to the bit width of the
$index$ because only the $index$ field is represented by qubits.
Thus the spatial complexity is $n$. As a comparison, since both the
fields $value$ and $target$ are represented by qubits, the spatial
complexity of the complete approach of Grover's
algorithm\cite{Kim_2002} is $3n$. Although the MSAS algorithm also
have a spatial complexity of $n$, it is not a complete database
search. Thus our algorithm has the best spatial complexity in the
quantum algorithms for complete database searches.

To evaluate the time complexity of our algorithm, a decisive
mathematical analysis is not possible. Therefore in this letter, we
use both the perturbative method\cite{Amin_localminima} and
numerical simulation\cite{Farhi} to examine the situation of the
time complexity.

In the perturbative approach\cite{Amin_localminima}, the time cost
of a adiabatic algorithm by either \emph{global} or \emph{local}
evolution can be written as
\begin{eqnarray}
\label{Tlg} T_{local}&\propto& \sqrt{T_{global}} \propto
\sqrt{|S^{-}|/|S^{+}|}, \\
\label{s2} S^{+} &\approx& \{z: h(z,f)  <
m_c \}, m_c \propto \frac{\log{1 /
\delta}}{\log \zeta^{+}}\nonumber\\
S^{-} &\approx& \{z: E_z < E_c \}, E_c \propto \frac{\log
{1/\zeta^{-}}}{\log{1 / \delta}},
\end{eqnarray}
where $S^{+}$ is a set containing the eigenstates of the problem
Hamiltonian which have a small Hamming distance towards the solution
state $|f\rangle$, while $S^{-}$ contains the ones which have low
energy levels. $|S|$ is the cardinality of set $S$. $\zeta^{\pm}$
are dimensionless parameters which are defined as
$\zeta^{\pm}\equiv\zeta(s^*\pm\epsilon_0)$. Here
$\zeta(t)=\frac{s(t)}{1-s(t)}$ and $s^*$ is the position of the
minimum gap between the ground and first exited state. $\epsilon_0$
and $\delta$ are small numbers.

To apply this result to our algorithm, we assume that the minimum
gap is on the central position of $s$, thus we can get $\zeta^{+}
=1/\zeta^{-}>1$. Then we define a small number
$\Omega\equiv\frac{\log{1 / \delta}}{\log \zeta^{+}}$. Because the
degeneracy of energy levels in the problem Hamiltonian in
Eq.\eqref{hp_hd} is $C_{n}^{i}$ where $i$ is the $i$th energy level,
Eq.\eqref{s2} goes as
\begin{eqnarray}
|S^{+}| &\approx& \sum_{i=0}^{i<m_c}{C_n^i}, m_c \propto \Omega\nonumber\\
|S^{-}| &\approx& \sum_{i=0}^{i<E_c}{C_n^i}, E_c \propto 1/\Omega.
\end{eqnarray}
Here, $\Omega$ is a small number and is not supposed to increase
with $n$, so only some low energy levels will be in $S^{-}$ and
$|S^{+}| $ is a comparatively small positive integer. Since
$\sum_{i=0}^{n-1}{C_n^i}= N$, for the worst case, we can take
Eq.\eqref{Tlg} as $T_{global} \propto|S^{-}|\propto N^\alpha$ where
$\alpha < 1$ is a constant.

\begin{figure}[t]
\begin{center}
\includegraphics[width= 1\columnwidth]{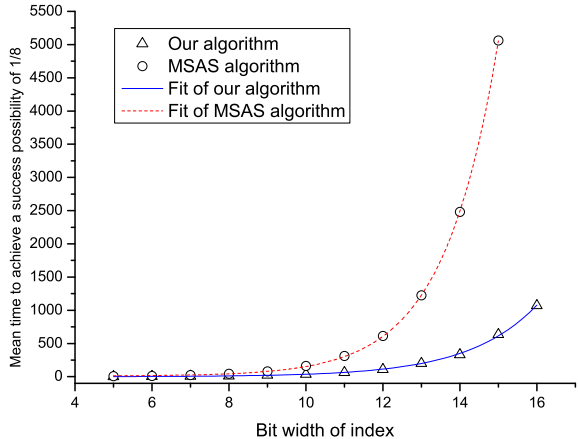}
\end{center}
\caption{Comparison of running time between our algorithm and MSAS
algorithm to achieve a success probability of 1/8 as the function of
bit width of $index$. The solid line is the fit of the triangles
each of which represent the mean simulation result of 50 instances
($32$ instances for $n=5$) of our algorithm, while the dash line is
the fit of the circles each of which is that of the MSAS algorithm.
The error range of fitting the triangles is $\pm2\%$ and that for
the circles is $\pm0.3\%$ } \label{time}
\end{figure}

To derive a more accurate range for $\alpha$, we performed a
numerical simulation\cite{Farhi} for randomly generated databases
with the bit width of $index$ sized from $5$ to $16$. For each bit,
we randomly generated 50 instances of database search. Then we
performed a numerical global evolution using four-order self-adapted
Runge-Kutta method to get a success probability of range
$[0.12,0.13]$ for each instance. The mean time for each bit is shown
in Fig.\ref{time}. By fitting the mean time, we obtain
$\alpha=0.81$. For a comparison, we simulated the time complexity of
MSAS algorithm using the same environment. The value $\alpha$ of
MSAS algorithm obtained from the fitting is $1.02$. The result of
simulation fit well with theoretical expectation where $\alpha$ is
$1$\cite{Farhi_2000}.

In Fig.\ref{time}, the running time of our algorithm grows much more
slowly than the MSAS algorithm. This result matches well with the
expectation from the perturbative analysis. Both the results show
that our algorithm has a better performance in time complexity than
the MSAS algorithm. And because local evolution can provide a
quadratic speedup over global evolution, theoretically the time
complexity of our algorithm by local evolution can be reduced to
less than order $\sqrt{N}$, even lower than than the complexity of
Grover's algorithm.

To be concluded, we introduce a new algorithm for quantum search
problem by adiabatic evolution. We use another method to represent
the quantum database in this algorithm and it saves $\frac{2}{3}$ of
qubits than the complete approach of Grover's
algorithm\cite{Kim_2002}. We use both the emerging perturbative
method of adiabatic algorithm and numerical simulation to analyze
the time complexity in this algorithm. The results show that it
provides a higher speedup than the MSAS algorithm and potentially
has a better performance than Grover's algorithm. This algorithm can
be experimentally verified in NMR or ion-trap
systems\cite{Peng_2008,Friedenauer_2008}.

The authors thank Zeyang Liao, Dieter Suter and Guilu Long for
discussions and comments. This work was supported by National Nature
Science Foundation of China, the CAS, Ministry of Education of PRC.

\end{document}